\documentclass[article]{aa}       

\usepackage{graphics}
\usepackage{epsfig}
\usepackage{latexsym}
\begin{document}
\hyphenation{brems-strah-lung}
\title{Outbursts from IGR J17473-2721}

\author{Shu Zhang \inst{1}, Yu-peng Chen \inst{1}, Jian-Min Wang \inst{1,2}, Diego F. Torres \inst{3}, Ti-Pei Li \inst{1,4}
}

\institute{Laboratory for Particle Astrophysics, Institute of High
Energy Physics, Beijing 100049, China
\and
Theoretical Physics Center for Science Facilities (TPCSF), CAS.
\and
ICREA \& Institut de Ci\`encies de l'Espai (IEEC-CSIC), Campus
 UAB, Facultat de Ci\`encies, Torre C5-parell, 2a planta, 08193 Barcelona, Spain 
              \and
Center for Astrophysics,Tsinghua University, Beijing 100084, China
         }

\offprints{Shu Zhang}
\mail{szhang@mail.ihep.ac.cn}

\date{Received  / Accepted }

\titlerunning{Outbursts from...}
\authorrunning{Shu Zhang et al.}

  \abstract
  {}
{IGR~J17473-2721 was  discovered by \emph{INTEGRAL} as  a neutron star low mass X-ray binary. To date, two outbursts have been detected in X-rays  by \emph{RXTE}/ASM and \emph{SWIFT}: the one occurring in 2005 was weak and was characterized by a low/hard state spectrum; the one of March 2008 was strong and showed a 6-step evolution in its flux. We investigate their  evolution,  emphasizing  the later outburst.  }
{We analyzed all available observations  carried out by \emph{RXTE} on  IGR J17473-2721 during its later outburst.
We analyzed as well all the available \emph{SWIFT}/BAT data on this source.
}
{The flux of the latter outburst rose in $\sim$ one month and then kept roughly constant for the following $\sim$ two months. During this time period, the source was in a low/hard state. The source moved to a high/soft state  within the following three days, accompanied by the occurrence of an additional outburst at soft X-rays and the end of the preceding outburst  in hard X-rays. During the decay of this soft outburst, the source went back to a low/hard state within 6 days, with a luminosity 4 times lower than the first transition. This shows a full cycle of the hysteresis in transition between the hard and the soft states.  The fact that the flux remained roughly constant for $\sim$ two months at times prior to the spectral transition to a high/soft state might be regarded as the result of balancing the evaporation of the inner disk and the inward accretion flow, in a  model in which the state transition is determined by  the mass flow rate.   Such a balance might be broken via an additional mass flow accreting onto the inner disk, which lightens the extra soft outburst  and causes the state transition.  However, the possibility of an origin of the emission from the jet during this time period cannot be excluded. The spectral analysis suggests an inclined XRB system for IGR J17473-2721. Such a long-lived preceding low/hard state makes IGR J17473-2721 resemble the behavior of outbursts seen in black hole X-ray binaries like GX 339-4.
} 
{} 


   \maketitle

\section{Introduction}

 X-ray binaries (XRBs) are generally classified as black hole (BH) or neutron systems according to the type of the compact star. The latter class is further grouped into the so-called Z and atoll sources, according to their behavior in spectral and time evolution. Z sources are usually brighter than the atoll ones, due to a higher accretion rate. 
The typical luminosity of atoll sources is generally less than 0.1 $\times$ the Eddington limit, while for Z sources it is $>$ 0.5 $\times$ the Eddington limit (Hasinger and van der Klis 1989). Atoll sources are mostly studied at their outbursts, during which the spectrum evolves from the {\sl island state} to the {\sl banana state}, which is analogous to the BH counterparts: the hard and the soft states, respectively. 
The spectral evolution of both types of  XRBs can be quite similar, as observed by \emph{RXTE} and \emph{BeppoSAX}. Those outbursts that (1) stay only in the hard state; (2) have a hard  outburst preceding a soft one; (3) show hysteresis in the transition between the hard and the soft states are particularly interesting.

 Meyer et al. (2000) believed  that the change in spectral state is related to the  modifications in the accretion flow and its balance  by the evaporation at the inner disk.  At the start  of the outburst, the accretion rate is low and the disk is cold and optically thick. The accretion flow in the inner disk region evaporates to form hot plasma or an ADAF region, to produce hard X-rays via Comptonizing the seed photons from the surface of the neutron star or the boundary layer. 
The sources stay in  the low/hard state. An increase in accretion flow toward the compact star might allow the hot plasma to be cooled and form an inner disk. As a result, the ADAF region is suppressed and the source moves into the high/soft state.  Mayer-Hofmeister et al. (2005) argued that the hysteresis could be a natural result of the cooling/heating of the corona by the photons produced at different mass accretion rates. 
Since the last decade, the outbursts of low mass X-ray binaries (LMXBs) have been  observed mainly by the high sensitivity detectors on board the satellites \emph{RXTE} and \emph{BeppoSAX} in  soft and hard X-rays. 
However, since the neutron star has a solid surface and magnetic field, spectral modeling the outburst in neutron star LMXBs could be more complicated than those with a BH. So far, two main paradigms exist to model outbursts: the Eastern model (Mitsuda et al. 1989) and the Western model (White et al. 1988). They differ in how the thermal and Comptonized components are dealt with. The Eastern model considers the thermal emission from the disk and the Comptonization from hot plasma around the neutron star, while the Western model considers the thermal emission from the boundary layer and the Comptonization from the disk. Although Done et al. (2007) showed that the fast variability of the hard component  of the energy spectrum could be  evidence against the Western model, in practice the models are usually hard to unambiguously resolve from the observational data, e.g. the existence of the so-called degeneracy in models. For example,  the hard component of the outbursts as observed in  NS LMXBs Aql X-1 and 4U 1608-52 by \emph{RXTE}  is   well described by a broken power law model (Lin et al. 2007), while  during the 1998 outburst of 4U 1608-52 is described by thermal Comptonization (Gierlinski et al. 2001).                 

 Since the satellites  like \emph{INTEGRAL} and \emph{SWIFT} began to produce data in 2002 and 2004, more atoll sources, for examples, IGR  J00291+5934 (Eckert et al. 2004; Linares 2008a) and SWIFT J1756.9-2508 (Krimm et al. 2007; Linares et al. 2008b), have been discovered in hard X-rays.  The hard X-ray source IGR J17473-2721 was discovered as an additional member of the sample by the IBIS/ISGRI telescope on board of INTEGRAL during the ultra deep Open Program observations of the 
Galactic center carried out on March 17-28 and April 9-21, 2005 (Grebenev et al. 2005).  
Following this discovery,  observations were carried out at other wavelengths as well, i.e. at soft X-rays by \emph{Chandra} with an exposure of 1.1ks, and in the infrared at  the Ks band by the \emph{PANIC} camera of the 6.5 m Magellan I telescope with an exposure of 270 seconds. 

 Historically, the first outburst from  IGR~J17473-2721  was observed in 2005, with a  flux level of $\sim$ 7.77$\pm$0.10$\times$10$^{-11}$ erg/cm$^2$/s at 0.5-10 keV, and a location at (Ra,Dec)=(17 47 18.06, -27 20 38.9) (Juett et al. 2005).  The second outburst was observed  three years later  by \emph{SuperAGILE}  at 17-25 keV  on March 26, 2008 (Del Monte et al. 2008). According to a joint analysis of  the intensity and the temporal and spectral properties, IGR~J17473-2721 was recognized as a type-I burster. This was later confirmed by a  \emph{SWIFT} observation carried out on March 31 2008, with an exposure of 4.1 ks. A 100-second type-I burst was detected  at 0.01-100 keV with a temperature  $\sim$ 2.29$^{+0.22}_{-0.18}$ keV and a bolometric flux $\sim$ 1.1$\times$10$^{-7}$ erg/cm$^2$/s (Altamirano et al. 2008a). From \emph{SWIFT} observations, a column density of $\sim$ $N_{\rm H}$=3.8$^{+0.48}_{-0.14}$$\times$10$^{22}$ cm$^{-2}$  and a spectral index of 1.68$^{+0.21}_{-0.07}$ were derived (Altamirano et al. 2008a). The source was estimated to be at a distance of $\sim$   3.9-5.4 kpc and to have an upper limit of  luminosity of $\sim$ 1$\times$10$^{36}$ erg/s (Altamirano et al. 2008a). The following observations were performed by \emph{INTEGRAL} and \emph{SWIFT} in hard X-rays on April 1 and 8, 2008. These observations show clearly that a very large burst occurred in the source (Kuulkers et al. 2008; Baldovin et al. 2008). Noticeably, the \emph{INTEGRAL}/JEMX detected another 50-second type-I burst of a peak flux $\sim$ 1.5 Crab and a black body temperature $\sim$ 2.4$\pm$0.4 keV (Baldovin et al. 2008).  The most recent \emph{RXTE} results identified the source as an atoll source (Altamirano et al. 2008b).

 Since both outburst events have been well monitored by \emph{SWIFT} in hard X-rays and by \emph{RXTE} in soft X-rays, we investigate the details of the outburst in these two neighboring X-ray bands.   We aim to investigate the evolution of the flux in hard and soft X-rays during the most recent outburst.

\section{Observations and Results}
\subsection{Lightcurves}
ASM is one of the three detectors on-board the \emph{RXTE}
satellite (Gruber et al. 1996), which  has been used to track the
long-term behavior of the source in the energy band 1.5-12 keV since
February 1996. The target source was usually observed several times
per day, with so-called dwells of 96 seconds duration  each.
The extracted source lightcurves are available at the energy bands 
of 1.5--3, 3--5, 5--12, and
1.5--12 keV. For IGR~J17473-2721, we take  the daily average at the energy band 1.5-12 keV (see Fig. 1). The ASM lightcurve shows that two outbursts occurred from the source.  The first one (Fig. 2) lasted for 95 days ($\sim$ MJD 53495-53590) and reached a peak flux of 6 ct/s ($\sim$ 80 mcrab). The flux rose in the first 35 days and then decayed in the following 60 days. The source became active again three years later, with  a start time at around MJD 54555. This huge outburst showed a more complex, 6-step  evolution in flux, and accordingly six time intervals are defined in Fig.  2. The flux  rose for $\sim$ one month (interval I) and then kept roughly constant for the next $\sim$ two months (interval II). In the following three days, the flux in soft X-rays  suddenly jumped to $\sim$ 30 ct/s ($\sim$ 400 mcrab, interval III) and then decayed (intervals IV and V). 

\emph{SWIFT} carries the Burst Alert Telescope (BAT, Barthelmy et al. 2005), which has a rather large field of view of $1.4$~sr in partially-coded mode and works in the $15-150$~keV energy band. This makes it possible for a source to be  monitored daily in  hard X-rays. The data products are therefore   lightcurves in the 15-50~keV energy band   and are publicly available\footnote{See the \emph{SWIFT}/BAT transient monitor results provided by the \emph{SWIFT} Team at \texttt{http://swift.gsfc.nasa.gov/docs/swift/results/transients/}}. The BAT lightcurve traces IGR~J17473-2721 back to February 12, 2005 (see Fig. 1), and covers the two  outbursts as observed by ASM. In hard X-rays of 15-50~keV, the  first outburst has a similar profile to that seen in the soft X-rays (Fig. 2). The peak flux is about 0.035 ct/cm$^2$/s ($\sim$ 153 mcrab). The second outburst has  a profile resembling that in  soft X-rays   until $\sim$ MJD 54634 (Fig. 2, intervals I and II). Later, while in soft X-rays the source had an additional huge  outburst, in hard X-rays the preceding  outburst had almost ceased. The flux dropped in  hard X-rays within three days from $\sim$ 0.075 ct/cm$^2$/s ($\sim$ 330 mcrab) to $\sim$ 0.01 ct/cm$^2$/s ($\sim$ 44 mcrab) (interval III). The flux remained at that low level (interval IV) during the decay of the  soft X-ray  outburst until MJD 54670, when a new  outburst in hard X-rays formed within 6 days (interval V). In the time interval VI (Fig. 2), this new  outburst has an average flux $\sim$ 0.0188 ct/cm$^2$/s at 15-50 keV, and $\sim$ 2.89 ct/s at 1.5-12 keV. The duration of the second outburst is longer than 140 days. 

\subsection{Hardness ratios}
The hardness ratios of the two  outbursts are constructed with a flux ratio
15-50 keV/1.5-12 keV, and are shown in Fig. 3. It is obvious that the spectrum remained  hard (e.g. in the low/hard state) for the whole first outburst period and as well for the first $\sim$ 80 days of the second outburst (intervals I and II). We performed a linear fit to the hardness ratio and show this in Fig. 3 for both the fit and the residual. It seems that for the first outburst the spectrum slightly  softens along the outburst evolution, although the data quality is not high enough to be certain of this.  For the second outburst, the spectrum becomes soft within three days, with the emission dominated by the soft X-rays (interval III). Accompanying this change is an additional sudden  outburst in soft X-rays and the cessation of the preceding  outburst in hard X-rays (interval III). Such a feature  is typical for LMXB in their spectral transitions from a low/hard to a high/soft state. Along with the decay in soft X-rays, the source stayed in a high/soft spectral state (interval IV) until the transition  in interval V. Within 6 days, the source had returned to a low/hard state at a relatively low luminosity level (interval VI). This shows a full cycle of the hysteresis in the transition between the hard and the soft states. 

\subsection{Spectra}

 During the long outburst of IGR J17473-2721,  93 \emph{RXTE}/PCA observations were made, with the identifier (OBSID) of proposal number (PN) 93442, available  in the data archive. These observations contain $\sim$ 225 ks of exposure on the source, and are scattered over the whole evolution of the outburst (see  Fig.  2). These observations smoothly cover the last 5 time intervals II-VI.
The analysis of PCA data is performed  using 
Heasoft v. 6.2.  We filter the data using the standard \emph{RXTE}/PCA criteria.
Only the top layer of PCU2 (in the 0-4 numbering scheme) has been used for the analysis, because only the PCU2 was 100\% on  during the observation,
and we use the time intervals with the following constraints:  elevation angle $>10^{\circ}$, and pointing offset $<0.02^{\circ}$. Elevation is the angle above the limb of the Earth and the pointing offset is the angular distance between the pointing and the source. The background file used in the analysis of PCA data is the most recent one  available from the HEASARC website for bright sources,
pca$\_$bkgd$\_$cmbrightvle$\_$eMv20051128.mdl. Data from cluster 1 of the HEXTE system are used to produce the spectra.
An additional  1\% systematic error is added to the
spectra because of calibration uncertainties, if not otherwise specified.  The
spectra are fitted with XSPEC v12.3.1 and the model parameters are
estimated  with a 90$\%$ confidence level. 

We investigate the spectrum by fitting the data from each of the 5 time intervals as shown in Fig.  2. For the thermal component, we used the multicolor disk black body  model (diskbb) and the black body model (bbodyrad), where the soft X-rays are supposed to come from the disk and the boundary layer or surface of the neutron star, respectively. For the hard component, we performed  trials with the thermal Comptonization model (comptt), the cutoff power law model (cutoffpl), and the broken power law model (bkn). For all the trials of the models, the central energy of the iron line is fixed at 6.4 keV, and the absorption is fixed at 3.8$\times$10$^{22}$ atoms/cm$^2$ (Altamirano et al. 2008a). The energy band is adopted at 3-30 keV for PCA and at 30-100 keV for HEXTE. A constant is added into the model to account for the different normalization between PCA and HEXTE. 

The normalization of the disk black body is described as 
$(R_{in}^2/D_{10}^2)cos(\theta)$, 
where the  R$_{in}$ in units of km, D$_{10}$ the source distance in units of 10 kpc, and cos($\theta$) the disk inclination. Generally, for XRB systems, an observation of an eclipse means that the accretion disk is edge-on, and therefore puts stronger constraints on the disk inclination.  Since for IGR J17473-2721  no eclipses are observed, we can constrain the normalization factor in the disk black body model.   If we take the companion star to be of solar size ($\sim$ 2.3 light seconds) and the semi-major axis of the orbit to be 50 light seconds, we have an upper limit of the inclination angle of tg(50./2.3), which gives  $\sim$ 87.4$^{\circ}$ in order to avoid having an eclipse in the lightcurves. By putting the source  as D$_{10}$ $\sim$ 0.39-0.54, and taking R$_{in}$ as greater than 10 km, we have the lower limit of the normalization of the disk black body as 15-30 or even larger.   In our model fitting we remove those where  the normalization of the disk black body is well below 30. Also, we do not consider   models  with a reduced $\chi^2$ larger than 1.2.

We find that for the time interval II  only  the model consisting of the black body and the  cutoff power law can  fit  the data well (see  Fig.  \ref{spectra} for the spectrum and Table 1 for the parameters). The reduced $\chi^2$ is 1.17 with 71 dofs. Replacing  the black body by a disk black body results in the normalization of $\sim$ 8 for the disk black body.  For all the other trials the reduced $\chi^2$ is too large to be acceptable ($\sim$ 2.1). For the time intervals III and IV, the source was too weak to be detected by HEXTE, and only PCA data are used in spectral fittings. We find that all the trials with the component disk black body suffer from the small normalization.  The models with a  broken power law and  black body can well fit the data in time intervals III and IV (with the reduced $\chi^2$ $\sim$ 0.6-0.7), but with the iron line missing. The model with a black body and thermal Comptonization can  fit the data in time interval III (with the reduced $\chi^2$ $\sim$ 0.8), but  the normalization parameter of the latter component cannot be well constrained. In summary, the best models for time intervals III (reduced $\chi^2$ $\sim$ 0.8) and IV (reduced $\chi^2$ $\sim$ 1.1) are those with the components of a black body and a cutoff power law (see  Fig.  \ref{spectra} for the spectral fitting and  Table 1 for the parameters). For the time interval V we  do not need the soft component of either a black body or a disk black body; a model with a broken power law  can  fit the data well. The reduced  $\chi^2$ is about 1.01 with 72 dofs (see  Fig.  \ref{spectra} for the spectrum and Table 1 for the parameters). In the other trials, the reduced $\chi^2$ values are acceptable but there are unconstrained parameters. For the time interval VI, there are three possible models: a black body plus broken power law ($\chi^2$ $\sim$ 1.03), thermal Comptonization plus black body ($\chi^2$ $\sim$ 1.07) and  thermal Comptonization plus disk black body ($\chi^2$ $\sim$ 0.91). The last one has the temperature of the inner disk around 1.3 keV, which should  in general be less than 1 keV, and the normalization of the disk black body around 70, which would be unlikely if the inner disk radius is  tens of km away from  the surface of the neutron star during the low/hard state of the decay. Therefore we present in Table 1  the parameters obtained from the first two models and in  Fig.  \ref{spectra} the spectral fittings.   We find in  Table 1 that, the iron line has  width roughly constant along  the outburst, while the normalization shows that the strength of the iron line is correlated to the emission level of the soft X-rays: in the time intervals III and IV, the strength of the iron line is about a factor of 2-4 higher than in the other three time intervals. 

\subsection{CCD and luminosity evolution}

The flux ratio of 4-6.4 keV/3-4 keV is defined as the soft color
whereas the ratio 9.7-16 keV/6.4-9.7 keV is defined as the hard color.
Using them, we produce  the color-color diagram (CCD) in  Fig.
\ref{CCD}. Such a color definition is consistent with that adopted
in Gladstone et al. (2007). The outburst evolution is clearly seen in
this diagram; it has a typical profile of an atoll source. During the
hard state before the additional soft  outburst, the source remains on the
island, and then moves to the banana branch in the following soft
state; after the transition to the hard state again, the source returns to
the island branch, showing a further feature unlike those 
generally seen in atoll sources. Along the evolution of the
outbursts, most atoll sources have  CCDs moving from the island
branch  to the lower left part of the banana branch and then to
the right part of the banana branch (Done et al. 2007; Hasinger $\&$
van der Klis 1989).

To further investigate  the CCD, we follow the analysis 
in Gladstone et al. (2007) on the \emph{RXTE} atoll samples. We perform 
spectral fittings to  the intrinsic soft and hard colors, which
are free from  absorption and independent of the instrument
response. The resulted CCD and the color-luminosity diagrams are
shown in  Fig. \ref{color}, which supports that  IGR J17473-2721
should belong to the diagonal group as defined in Gladstone et al.
(2007) for the atoll source classification. 
 During outburst decay back to the hard state, the trace in the color-color diagram is vertical at a luminosity around 0.03-0.05 L$_{Edd}$ (for distances of 3.9-5.4 kpc).
 This occurs for the first atoll source of the diagonal group,
with the full cycle of  hysteresis  observed during the outburst.

\section{Discussion }

In the outbursts of IGR J17473-2721, we have seen almost the entire range of  state evolution so far detected in other atoll sources: the preceding hard state, the soft state and the  hysteresis in state transitions between the hard and the soft. 
 By comparing the outburst evolution between hard and  soft X-rays for the newly discovered neutron star LMXB system IGR~J17473-2721, we have found some other interesting behavior. The first of its outbursts was rather weak and the source  stayed in the low/hard state. So did the second outburst for the time period prior to the first state transition. This  resembles the small subset of  LMXB  transients like GS 1354-64, GRO J0422+32, GRS 1719 and XTE J1118+480, which were observed during an outburst to have  a spectrum remaining in a low/hard state with no transition to a high/soft state  (e.g. Brocksopp et al. 2001). 
An explanation of this attributes the spectral transition to the change in accretion mass rate  (Meyer et al. 2000).   Along with the increasing  mass flow rate, the evaporation and inward mass flow become dynamically balanced and, as a result, the truncated inner disk moves  closer to the compact object. Such a procedure is responsible for the luminosity rise  of the outburst, and the source stays for most of the time in a low/hard  state.  Given  an individual LMXB system, when the mass flow rate exceeds the critical value,  the balance between evaporation and inward flow will be  destroyed. A direct result of this is the ceasing of the outburst in hard X-rays and increasing  emission dominated by the soft X-rays, e.g. to form a so-called high/soft state. Otherwise the source  stays in a low/hard state for the whole life of the outburst  (Meyer-Hofmeister 2004).   

 For IGR~J17473-2721, instead of forming a sharp peak at the end of the flux rise, the  flux does not change much for the next two months in a low/hard state. This might be regarded as evidence for  a steady mass flow moving inward and  an ADAF filled inner region  remaining for quite a long time.  The mass flow rate did not increase during this time period to reach the critical rate for the spectral transition.  An even more interesting feature is that, after the long-lived low/hard state, the source finally moved to a high/soft state. Such a transition  requires the balance to be upset  via  additional mass flow accreting into the inner region.                 

 We notice that the outburst of IGR~J17473-2721 in hard X-rays is a long-live plateau rather than a sharp peak,  typical of the previous samples like Aql X-1. The entire outburst evolution of IGR J17473-2721  resembles the 1998 outburst of the BH  XRB  GX 339-4.   GX 339-4 stayed in a hard state for 500 days prior to transition to a soft state within roughly 100 days.  The entire outburst lasted for at least 400 days, if the start time is the transition to soft state (Yu et al. 2007, Belloni et al. 1999). The outburst of IGR J17473-2721 is more similar to GX339-4 based on  their long-lived preceding outburst in hard X-rays. To date, Aql X-1 is among the few NS to have an outburst similar to GX339-4 by showing the hysteresis, but the obvious difference of forming a sharp peak in the preceding hard X-ray outburst limits its similarity.  Therefore, the long-lived preceding hard outburst and the clearly established hysteresis with the luminosity differing by a factor of four between the state transitions, make IGR J17473-2721  resemble  the BH XRBs  in  outburst more than the other few known atolls. 

As it is currently known, neutron star LMXBs have spectral states comparable to those  in BH XRBs, when in outburst. The state transitions and the outburst evolution of both atolls  and BH XRBs are likely described in a similar scenario, where the accretion rate and the balance between inward accretion flow and the evaporation of the inner disk play an important role.
Although the spectral changes are obvious for LMXB in outburst, their modelling is more complicated than in BH XRBs due to the uncertainty attributed to the emission from the surface of the neutron star or its boundary layer.
The atoll sources Aql X-1 and 4U 1608-52 are examples: they have been observed so far (by \emph{RXTE}) in more than 20 outbursts. Lin et al. (2007) find that, instead of the classic  Comptonization model for the hard component, a broken power law model works better both for the hard and the soft states. 
For 4U 1608-52, its 1998 outburst was well studied as well by Gierlinski and Done (2001), who found that the non-thermal component of the spectrum can be well represented in a Comptonization plus Compton reflection model. 
These findings indicate the difficulty in  unambiguously  modeling the spectrum of the atolls  when in outburst. 
For IGR J17473-2721 the spectrum of the high/soft state can be well represented  using the model consisting of a black body and cutoff power law.  That the hard component for most of the outburst can be fitted by the non-Comptonization model in its the banana branch is consistent with the spectral fits of Aql X-1 and 4U 1608-52 in outburst  using a broken power law model as the hard component (Lin et al. 2007). The  broken power law component can be considered as Comptonization under complex conditions or in combination with other radiation process (Lin et al. 2007). We notice that the data for the long-lived hard state preceding to the hard/soft transition can only be fitted by using a  model with the components of the black body and the cutoff power law. Unlike  the other time intervals (III-VI) where a spectral degeneracy exists, the thermal Comptonization model can be simply rejected from the obviously unacceptable $\chi^2$. This may provide  evidence that the hard X-rays in this long-lived hard state may not be the Comptonization emission and may instead originate in  a jet. We notice as well that for IGR J17473-2721 the disk black body model is likely not  appropriate for almost the whole evolution of the outburst; those normalizations of the black body used for fitting the soft component in the spectrum are relatively larger during the hard states than during the soft states. These features may be understood in a largely inclined XRB system: due to the large inclination of the disk, the projection of the disk emission area to  the line of sight becomes small and hardly visible in the energy spectrum. The soft component of the spectrum may come from the area near the surface of the neutron star, and be surrounded by the corona region of the hot plasma  which is optically thin in the hard state. 
 
In summary, the newly discovered neutron star LMXB IGR~J17473-2721 showed interesting  features in its outburst evolution, with a long-lived  plateau  hard  outburst followed by a spectral transition to the high/soft state and a further increase of luminosity during the high/soft state.
 In the model of a truncated accretion disk this might be interpreted as a consequence of   a steady mass flow moving inward, 
 a truncated inner disk and an ADAF filled inner region staying for a long time during the outburst evolution of IGR~J17473-2721.  
The possibility of an origin of the emission from the jet during this time period  also cannot be excluded. The spectral analysis  suggests an inclined XRB system for IGR J17473-2721.   The full cycle of  hysteresis together with 
 a long-lived plateau  in the outburst light curve during the hard state, preceding the hard/soft spectral transition, 
 make IGR~J17473-2721 more  analogous to the BH XRBs   than any other atolls observed previously.

\acknowledgements  
This work was subsidized by the National Natural Science Foundation of China, the CAS key Project KJCX2-YW-T03, and 973 program 2009CB824800. J.-M. W. thanks the Natural Science Foundation of China for support via NSFC-10325313, 10521001 and 10733010. DFT acknowledges support from grants AYA 2006-00530 and CSIC-PIE 200750I029.

\clearpage

\begin{figure*}[ptbptbptb]
\centering
 \includegraphics[angle=0, scale=0.7]{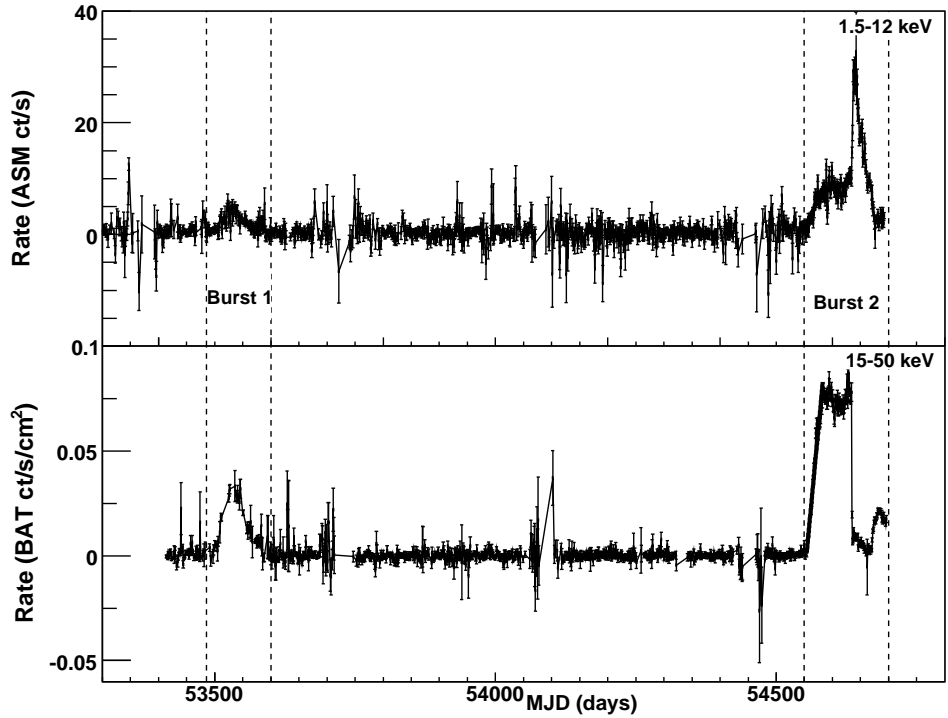}
      \caption{ASM lightcurve (1.5-12 keV, upper panel) and \emph{SWIFT}/BAT lightcurve (15-50 keV, lower panel) covering the two outbursts of IGR J17473-2721. The dashed lines show the start and the end of each outburst. }
         \label{lc-asm-swift}
\end{figure*}

\begin{figure*}[ptbptbptb]
\centering
 \includegraphics[angle=0, scale=0.7]{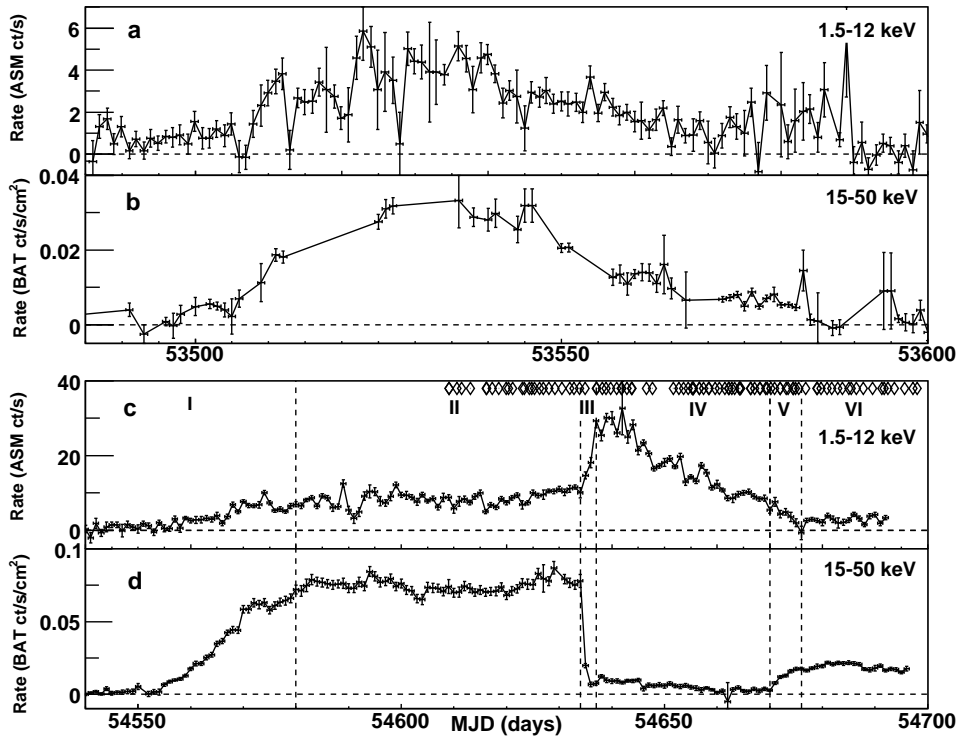}
      \caption{Flux evolution of the first outburst   as observed in ASM (panel a) and in \emph{SWIFT}/BAT (panel b), and of the second outburst as observed in ASM (panel c) and in \emph{SWIFT}/BAT (panel d). The dashed lines in panels c and d denote six steps in flux evolution. The available PCA observations are marked at  the top  the panel c (diamonds).}
         \label{1st-burst}
\end{figure*}

\clearpage

\begin{figure*}[ptbptbptb]
\centering
 \includegraphics[angle=0, scale=0.7]{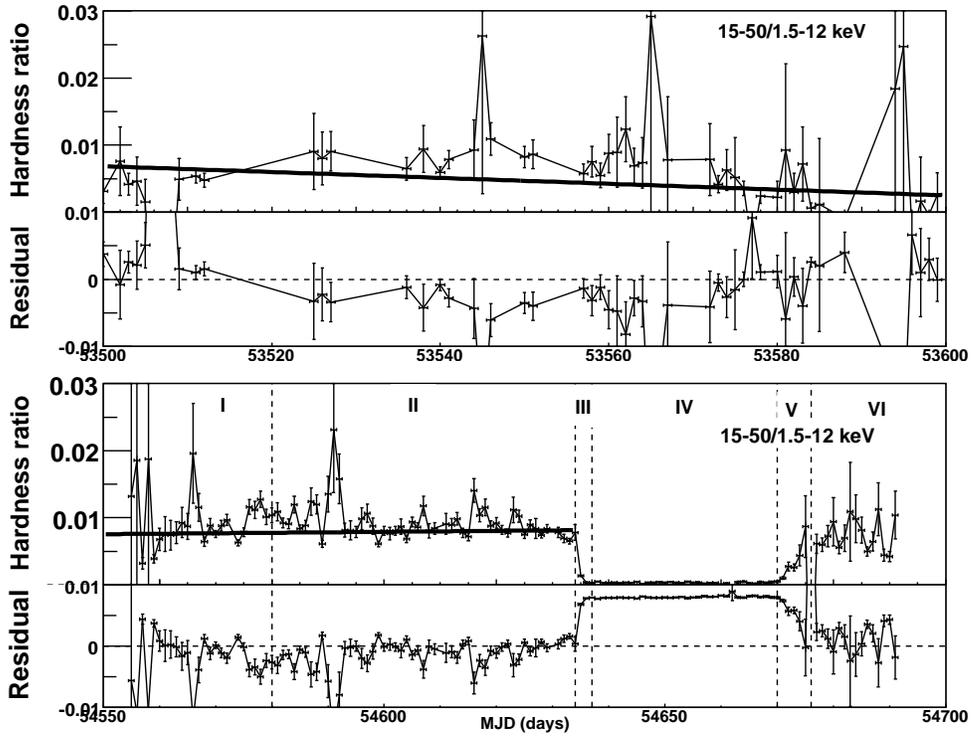}
      \caption{Evolution of the hardness ratio   (15-50 keV/1.5-12 keV)  of the first outburst  (upper panel) and of the second outburst (lower panel). The linear fits are shown by the solid lines and the residuals are given in  the lower part of each figure.  }
         \label{hardness-ratio}
\end{figure*}

\begin{figure*}[ptbptbptb]
\centering
 \includegraphics[angle=0, scale=0.7]{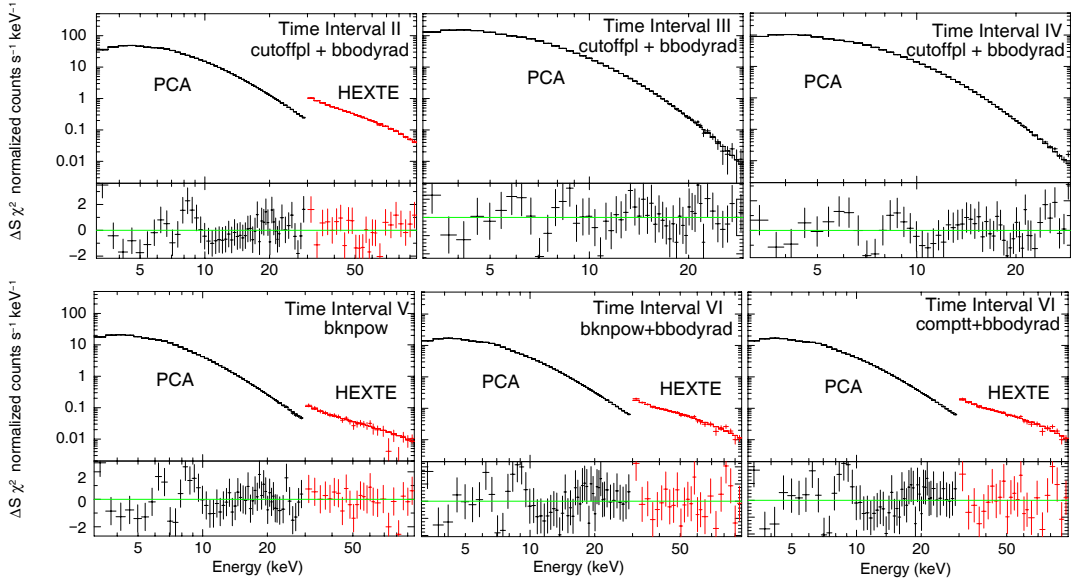}
      \caption{The spectra and the best fit models for each of the five time intervals of the outburst.}
         \label{spectra}
\end{figure*}

\clearpage

\begin{figure*}[ptbptbptb]
\centering
 \includegraphics[angle=0, scale=0.7]{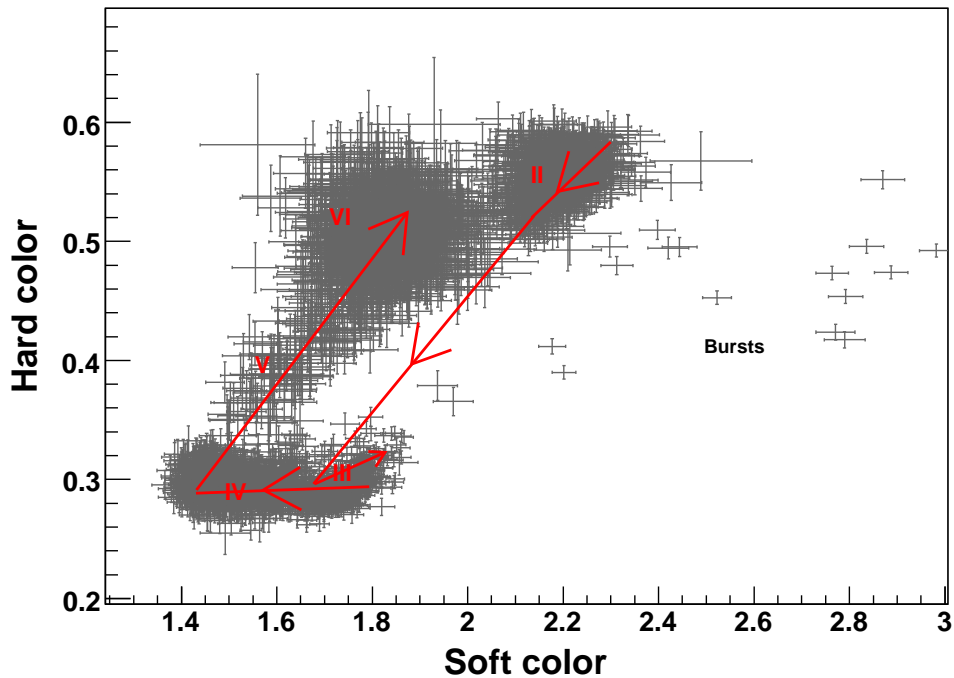}
      \caption{The CCD of IGR J17473-2721. The hard color  (9.7-16 keV/6.4-9.7 keV) and the soft color  (4-6.4 keV/3-4 keV) are calculated by using the PCA count rate. The lines show the track in flux evolution of the outburst. 16 Type I bursts are shown at the region with large values of soft color.}
         \label{CCD}
\end{figure*}


\begin{figure*}[ptbptbptb]
\centering
 \includegraphics[angle=0, scale=0.7]{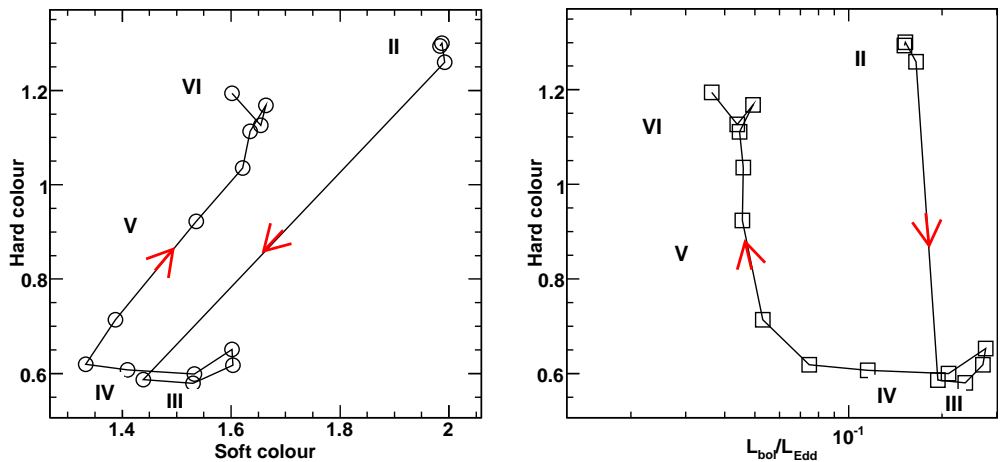}
      \caption{ The intrinsic CCD (left) and the color-luminosity (right) diagram of IGR J17473-2721. The hard color  (9.7-16 keV/6.4-9.7 keV) and the soft color  (4-6.4 keV/3-4 keV) are calculated by using the energy spectrum fit to the PCA/HEXTE data. The  luminosity is estimated at 3-30 keV at a distance of 5.4 kpc. The five time intervals are marked with the numbers II-VI. }
         \label{color}
\end{figure*}

\clearpage

\begin{table*}[ptbptbptb]
\begin{center}
\label{tab1}
\caption{The results from  fitting PCA/HEXTE data  with different models.}
\vspace{5pt}
\small
\begin{tabular}{cccccc}
\\\hline
Model&Parameters (units)&\multicolumn{3}{c}{Time Interval }   
  \\

  & &II &III    &IV \\\hline

bbodyrad + & $kT_{bb}$ (keV)          &$ 1.11_{-0.22}^{+0.07}$  & $2.84_{-2.84}^{+1.37}$     &$1.98_{-0.08}^{+0.05}$  \\
           & $N_{bb}$            &$38.29_{-5.89}^{+41.76}$  &$ 0.44_{-0.44}^{+1.70}$     &$ 4.97_{-1.65}^{+1.50}$  \\
 cutoffpl   & $\Gamma $           &$1.30_{-0.04}^{+0.04}$  &$1.45_{-0.43}^{+0.22}$     &$1.89_{-0.14}^{+0.02}$  \\
           &  $E_{cut}$ (keV)         &$ 38.51_{-1.90}^{+2.11}$  &$4.61_{-1.41}^{+0.65}$    &$6.37_{-0.68}^{+0.40}$ \\
           & $N_{cutoffpl}$      &$ 0.40_{-0.04}^{+0.04}$ &$4.65_{-0.60}^{+0.80}$    &$4.23_{-0.18}^{+0.13}$ \\
           &$\sigma_{Fe}$ (keV)       &0.6$^{+0.3}_{-0.2}$ & 0.7$\pm$0.2  &0.8$\pm$0.1 \\
           &$N_{Fe}$             &9$^{+7}_{-2}$  & 20$\pm$4  & 20$^{+2}_{-3}$\\
           & $\chi^{2}_{\nu}$/dof&1.17/71 & 0.83/47   &1.10/47\\\hline
Model&Parameters (units)&\multicolumn{3}{c}{Time Interval }   
  \\
  & & &V    &VI \\\hline
bbodyrad +     	   &$kT_{bb}$ (keV) && &$ 0.72_{-0.06}^{+0.08}$ \\
	   &$N_{bb}$&& &$77.32_{-32.72}^{+52.12}$\\
 bkn           &$\Gamma_{1} $ &&$ 2.26_{-0.01}^{+0.01}$  &$1.85_{-0.02}^{+0.02}$ \\
	   &$E_{bkn}$(keV)&&$12.15_{-1.15}^{+1.27}$&$ 50.98_{-7.43}^{+6.93}$  \\
           &$\Gamma_{2} $&&$2.14_{-0.04}^{+0.03}$ &$2.70_{-0.33}^{+0.46}$\\
           &$N_{bkn}$ &&$ 0.82_{-0.02}^{+0.02}$&$0.31_{-0.01}^{+0.01}$\\
	   &$\sigma_{Fe}$(keV)&& 0.8$\pm$0.1 &0.7$\pm$0.1\\
	   &$N_{Fe}$ &&5.0$\pm$0.5 &5.0$\pm$0.6\\
	   & $\chi^{2}_{\nu}$/dof&&1.01/72&1.03/70\\\hline
bbodyrad + &$kT_{bb}$ (keV) &&&$ 0.66_{-0.04}^{+0.04}$ \\
 	   &$N_{bb}$ &&&$330.02_{-70.52}^{+94.04}$\\
Comptt     &$kT_{seed}$ (keV)&&&$1.13_{-0.35}^{+0.11}$\\
	   &$kT_{e}$ (keV)&&&$21.83_{-4.14}^{+4.04}$\\
	&$\tau_{T}$&&&$1.87_{-0.23}^{+0.29}$\\
	&$N_{comp}$&&&$0.007_{-0.002}^{+0.001}$ \\
	&$\sigma_{Fe}$(keV)&&&0.7$\pm$0.1\\
	&$N_{Fe}$&&&5.0$\pm$0.5\\
	&$\chi^{2}_{\nu }$/dof &&&1.07/70 \\\hline
\end{tabular}
\end{center}
\begin{list}{}{}
\item[Note:]{All the fits have the iron line fixed at 6.4 keV and absorption at 3.8$\times$10$^{22}$atoms/cm$^2$. By putting the source at a distance of 5.4 kpc, the luminosities calculated at 3-30 keV are 19.7$\times$10$^{36}$ erg/s for time interval II,  24.9$\times$10$^{36}$ erg/s for time interval III,  17.6$\times$10$^{36}$ erg/s for time interval IV, 5.7 $\times$10$^{36}$ erg/s for time interval V, 5.6$\times$10$^{36}$ erg/s for time interval VI.  N$_{Fe}$ is the iron line normalization in units of 10$^{-3}$ ph cm$^{-2}$s$^{-1}$. The other normalizations $N_{cutoffpl}$ and $N_{bkn}$  are in units of cm$^{-2}$s$^{-1}$keV$^{-1}$. $N_{bb}$ is the black body normalization proportional to the surface area, defined as $R_{km}^{2}/D_{10}^{2}$, where $R_{km}$ is the surface radius in km and $D_{10}$ is the source distance in 10 kpc.}
\end{list}
\end{table*}
\clearpage

\end{document}